\documentclass[a4paper]{llncs}
\usepackage{amsmath,amssymb,bbm,hyperref}

\catcode`\à=\active \defà{\`a} \catcode`\À=\active \defÀ{\`A}
\catcode`\á=\active \defá{\'a} \catcode`\Á=\active \defÁ{\'A}
\catcode`\ä=\active \defä{\"a} \catcode`\Ä=\active \defÄ{\"A}
\catcode`\â=\active \defâ{\^a} \catcode`\Â=\active \defÂ{\^A}
\catcode`\å=\active \defå{{\aa}} \catcode`\Å=\active \defÅ{{\AA}}
\catcode`\ç=\active \defç{\c{c}} \catcode`\Ç=\active \defÇ{\c{C}}
\catcode`\è=\active \defè{\`e} \catcode`\È=\active \defÈ{\`E}
\catcode`\é=\active \defé{\'e} \catcode`\É=\active \defÉ{\'E}
\catcode`\ë=\active \defë{\"e} \catcode`\Ë=\active \defË{\"E}
\catcode`\ê=\active \defê{\^e} \catcode`\Ê=\active \defÊ{\^E}
\catcode`\ì=\active \defì{\`{\i}} \catcode`\Ì=\active \defÌ{\`{\I}}
\catcode`\í=\active \defí{\'{\i}} \catcode`\Í=\active \defÍ{\'{\I}}
\catcode`\ï=\active \defï{\"{\i}} \catcode`\Ï=\active \defÏ{\"{\I}}
\catcode`\î=\active \defî{\^{\i}} \catcode`\Î=\active \defÎ{\^{\I}}
\catcode`\ò=\active \defò{\`o} \catcode`\Ò=\active \defÒ{\`O}
\catcode`\ó=\active \defó{\'o} \catcode`\Ó=\active \defÓ{\'O}
\catcode`\ö=\active \defö{\"o} \catcode`\Ö=\active \defÖ{\"O}
\catcode`\ô=\active \defô{\^o} \catcode`\Ô=\active \defÔ{\^O}
\catcode`\ù=\active \defù{\`u} \catcode`\Ù=\active \defÙ{\`U}
\catcode`\ú=\active \defú{\'u} \catcode`\Ú=\active \defÚ{\'U}
\catcode`\ü=\active \defü{\"u} \catcode`\Ü=\active \defÜ{\"U}
\catcode`\û=\active \defû{\^u} \catcode`\Û=\active \defÛ{\^U}
\catcode`\ý=\active \defý{\'y} \catcode`\Ý=\active \defÝ{\'Y}
\catcode`\ÿ=\active \defÿ{\"y} \catcode`\˜=\active \def˜{\"Y}
\catcode`\½=\active \def½{!`}
\catcode`\¾=\active \def¾{?`}
\catcode`\ß=\active \defß{{\ss}}
\newcommand{\tmhlink}[2]{{#1}}
\newcommand{\tmabbr}[1]{#1}
\newcommand{\tmvar}[1]{\ensuremath{#1}}
\newcommand{\code}[1]{\texttt{#1}}
\newcommand{\tmmathbf}[1]{\ensuremath{\boldsymbol{#1}}}
\newcommand{\tmem}[1]{{\em #1\/}}
\newcommand{\tmop}[1]{\operatorname{#1}}
\newcommand{\nin}{\not\in}
\newcommand{\tmkbd}[1]{\texttt{#1}}
\newcommand{\defeq}{\stackrel{\mathrm{def}}{=}}
\begin{document}

\title{Essential Incompleteness of Arithmetic Verified by Coq} \author{Russell
O'Connor}
\institute{
Institute for Computing and Information Science\\
Faculty of Science\\
Radboud University Nijmegen\\
\and The Group in Logic and the Methodology of Science\\
University of California, Berkeley\\
\email{r.oconnor@cs.ru.nl}
\thanks{This paper appears in the proceedings of the 18th International
Conference on Theorem Proving in Higher Order Logics (TPHOLs 2005)}} \maketitle

\begin{abstract}
  A {\tmhlink{constructive proof of the Gödel-Rosser incompleteness
  theorem}{http://coq.inria.fr/contribs/GodelRosser.html}}
  {\cite{oconnor:2005}} has been completed
  using {\tmhlink{the Coq proof assistant}{http://coq.inria.fr/}}. Some theory
  of classical first-order logic over an arbitrary language is formalized. A
  development of primitive recursive functions is given, and all primitive
  recursive functions are proved to be representable in a weak axiom system.
  Formulas and proofs are encoded as natural numbers, and functions operating
  on these codes are proved to be primitive recursive. The weak axiom system
  is proved to be essentially incomplete. In particular, Peano arithmetic is
  proved to be consistent in {\tmhlink{Coq}{http://coq.inria.fr/}}'s type
  theory and therefore is incomplete.
\end{abstract}

\setcounter{section}{-1}
\section{License}

\begin{sloppypar}
This work is hereby released into the Public Domain. To view a copy of the
public domain dedication,
visit \url{http://creativecommons.org/licenses/publicdomain/} or send a letter
to Creative Commons, 559 Nathan Abbott Way, Stanford, California 94305,
USA.
\end{sloppypar}

\section{Introduction} \label{I}

The Gödel-Rosser
{\tmhlink{incompleteness theorem for arithmetic}{\# TSI}} states
that any complete first-order theory of a nice axiom system, using only the
symbols \tmmathbf{+}, \tmmathbf{{\times}}, \tmmathbf{0}, \tmmathbf{S}, and
\tmmathbf{<} is inconsistent. A nice axiom system must contain {\tmhlink{the nine
specific axioms of a system called NN}{\# AXIOMSYSTEMS}}. These nine axioms
serve to define the previous symbols. A nice axiom system must also be
{\tmhlink{expressible}{\# EXPRESSIBLE}} in itself. This last restriction
prevents the incompleteness theorem from applying to axioms systems such as
the true first order statements about $\mathbbm{N}$.

A computer verified proof of Gödel's incompleteness theorem is not new. In
1986 Shankar created a proof of the incompleteness of Z2, hereditarily finite
set theory, in the Boyer-Moore theorem prover {\cite{shankar:1994}}. My work
is the first computer verified proof of the essential incompleteness of
arithmetic. {\tmhlink{Harrison}{http://www.cl.cam.ac.uk/users/jrh/}}
recently completed a proof in HOL Light {\cite{harrison:2000}} of the
essential incompleteness of $\Sigma_1$-complete theories, but has not shown
that any particular theory is $\Sigma_1$-complete. His work will be included
in the next release of HOL Light.

My proof was developed and checked in {\tmhlink{Coq}{http://coq.inria.fr/}}
7.3.1 using {\tmhlink{Proof General}{http://proofgeneral.inf.ed.ac.uk/}} under
{\tmhlink{XEmacs}{http://www.xemacs.org/}}. It is {\tmhlink{part of the user
contributions to Coq}{http://coq.inria.fr/contribs/GodelRosser.html}} and can
be checked in {\tmhlink{Coq}{http://coq.inria.fr/}} 8.0 {\cite{CoqManualV8}}.
Examples of source code in this document use the new
{\tmhlink{Coq}{http://coq.inria.fr/}} 8.0 notation.

{\tmhlink{Coq}{http://coq.inria.fr/}} is an implementation of the calculus of
(co)inductive constructions. This dependent type theory has intensional equality
and is constructive, so my proof is constructive. Actually the proof depends
on the {\tmhlink{Ensembles
library}{http://coq.inria.fr/library/Coq.Sets.Ensembles.html}} which declares
an {\tmhlink{axiom of extensionality for
{\code{Ensembles}}}{http://coq.inria.fr/library/Coq.Sets.Ensembles.html \#
Extensionality\_Ensembles}}, but this axiom is never used.

This document points out some of the more interesting problems I encountered
when formalizing the incompleteness theorem. My proof mostly follows the
presentation of incompleteness given in
{\tmhlink{{\tmem{An Introduction to Mathematical
Logic}}}{URN:ISBN:053494440X}} {\cite{hodel:1995}}. I referred to the
{\tmhlink{supplementary text}{http://www.math.uwaterloo.ca/\~{
}snburris/htdocs/scav/fo\_arith/fo\_arith.html}} for the book
{\tmhlink{{\tmem{Logic for Mathematics and Computer
Science}}}{http://www.math.uwaterloo.ca/\~{ }snburris/htdocs/lmcs.html}}
{\cite{burris:1997}} to construct Gödel's $\beta$-function. I also use part of
Caprotti and Oostdijk's contribution of {\tmhlink{Pocklington's
criterion}{http://coq.inria.fr/contribs/pocklington.html}}
{\cite{oostdijk:2001}} to prove the Chinese remainder theorem.

This document is organized as follows. First I discuss the difficulties I had
when formalizing {\tmhlink{classical first-order logic}{\# FOCL}} over an
arbitrary language. This is followed by the definition of a language LNN and
an axiom system called NN. Next I give {\tmhlink{the statement of the
essential incompleteness of NN}{\# TSI}}. Then I briefly discuss
{\tmhlink{coding formulas and proofs as natural numbers}{\# C}}. Next I
discuss {\tmhlink{primitive recursive functions}{\# PRF}} and the problems I
encountered when trying to prove that substitution can be computed by a
primitive recursive function. Finally I briefly discuss the
{\tmhlink{fixed-point theorem, Rosser's incompleteness theorem}{\# FPTRIT}},
and the incompleteness of {\tmabbr{PA}}. At the end I give some 
remarks about how to extend my work in order to formalize
Gödel's second incompleteness theorem.

\subsection{{\tmhlink{Coq}{http://coq.inria.fr/}} Notation}

For those not familiar with {\tmhlink{Coq}{http://coq.inria.fr/}} syntax, here
is a short list of notation
\begin{itemize}
  \item {\code{->}}, \verb!/\!, \verb!\/!,
  and {\code{\~{ }}} are the logical connectives $\Rightarrow$, $\wedge$,
  $\vee$, and $\neg$.
  
  \item {\code{A -> B}}, {\code{A * B}}, and {\code{A + B}} form function
  types, Cartesian product types, and disjoint union types.
  
  \item {\code{*}}, {\code{+}}, and {\code{S}} are the arithmetic
  operations of multiplication, addition, and successor.
  
  \item {\code{inl}} and {\code{inr}} are the left and right injection
  functions of types {\code{A -> A + B}} and {\code{B -> A + B}}.
  
  \item {\code{::}}, and {\code{++}} are the list operations cons, and
  append.
  
  \item {\code{\_}} is an omitted parameter that
  {\tmhlink{Coq}{http://coq.inria.fr/}} can infer itself.
\end{itemize}

For more details see the {\tmhlink{Coq 8.0 reference
manual}{http://coq.inria.fr/doc/main.html}} {\cite{CoqManualV8}}.

\section{First-Order Classical Logic} \label{FOCL}

I began by developing the theory of first order classical logic inside
{\tmhlink{Coq}{http://coq.inria.fr/}}. In essence
{\tmhlink{Coq}{http://coq.inria.fr/}}'s logic is a formal metalogic to reason
about this internal logic.

\subsection{Definition of {\code{Language}}}

\begin{sloppypar}
I immediately took advantage of {\tmhlink{Coq}{http://coq.inria.fr/}}'s
dependent type system by defining {\code{Language}} to be a dependent record
of types for symbols and an arity function from symbols to $\mathbbm{N}$. The
{\tmhlink{Coq}{http://coq.inria.fr/}} code is:
\begin{verbatim}
Record Language : Type := language
  {Relations : Set; 
   Functions : Set; 
   arity : Relations + Functions -> nat}.
\end{verbatim}
In retrospect it would have been slightly more convenient to use two arity
functions instead of using the disjoint union type.
\end{sloppypar}

This approach differs from {\tmhlink{Harrison's definition of first order
terms and formulas in HOL
Light}{http://www.cl.cam.ac.uk/users/jrh/papers/model.html}}
{\cite{harrison:1998}} because {\tmhlink{HOL
Light}{http://www.cl.cam.ac.uk/users/jrh/hol-light/}} does not have dependent
types. Dependent types allow the type system to enforce that all terms and
formulas of a given language are well formed.

\subsection{Definition of {\code{Term}}}

For any given language, a {\code{Term}} is either a variable indexed by a
natural number or a function symbol plus a list of {\tmvar{n}} terms where
{\tmvar{n}} is the arity of the function symbol. My first attempt at writing
this in {\tmhlink{Coq}{http://coq.inria.fr/}} failed.
\begin{verbatim}
Variable L : Language.
(* Invalid definition *)
Inductive Term0 : Set :=
  | var0 : nat -> Term0
  | apply0 : forall (f : Functions L) (l : List Term0), 
       (arity L (inr _ f))=(length l) -> Term0.
\end{verbatim}
The type {\code{(arity L (inr \_ f))=(length l)}} fails to meet
{\tmhlink{Coq}{http://coq.inria.fr/}}'s {\tmhlink{positivity requirement for
inductive types}{http://coq.inria.fr/doc/Reference-Manual006.html \#
Positivity}}. Expanding the definition of {\code{length}} reveals a hidden
occurrence of {\code{Term0}} which is passed as an implicit argument to
{\code{length}}. It is this occurrence that violates the {\tmhlink{positivity
requirement}{http://coq.inria.fr/doc/Reference-Manual006.html \# Positivity}}.

My second attempt met the {\tmhlink{positivity
requirement}{http://coq.inria.fr/doc/Reference-Manual006.html \# Positivity}},
but it had other difficulties. A common way to create a polymorphic lists of
length {\code{n}} is:
\begin{verbatim}
Inductive Vector (A : Set) : nat -> Set :=
  | Vnil : Vector A 0
  | Vcons : forall (a : A) (n : nat), 
       Vector A n -> Vector A (S n).
\end{verbatim}
Using this I could have defined {\code{Term}} like:
\begin{verbatim}
Variable L : Language.

Inductive Term1 : Set :=
  | var1 : nat -> Term1
  | apply1 : forall f : Functions L, 
       (Vector Term1 (arity L (inr _ f))) -> Term1.
\end{verbatim}
My difficulty with this definition was that the induction principle generated
by {\tmhlink{Coq}{http://coq.inria.fr/}} is too weak to work with.

Instead I created two
mutually inductive types: {\code{Term}} and {\code{Terms}}.
\begin{verbatim}
Variable L : Language.

Inductive Term : Set :=
  | var : nat -> Term
  | apply : forall f : Functions L, 
       Terms (arity L (inr _ f)) -> Term
with Terms : nat -> Set :=
  | Tnil : Terms 0
  | Tcons : forall n : nat, 
       Term -> Terms n -> Terms (S n).
\end{verbatim}
Again the automatically generated induction principle is too weak, so I used
the {\code{Scheme}} command to generate suitable mutual-inductive principles.

The disadvantage of this approach is that useful lemmas about
{\code{Vector}}s must be reproved for {\code{Term}}s. Some of these lemmas
are quite tricky to prove because of the dependent type. For example, proving
{\code{forall x : Terms 0, Tnil = x}} is not easy.

Recently, Marche has shown me that the {\code{Term1}} definition would
be adequate. One can explicitly make a sufficient induction principle by using
{\tmhlink{nested {\code{Fixpoint}}
functions}{http://pauillac.inria.fr/pipermail/coq-club/2005/001641.html}}
{\cite{marche:2005}}.

\subsection{Definition of {\code{Formula}}}

The definition of {\code{Formula}} was straightforward.
\begin{verbatim}
Inductive Formula : Set :=
  | equal : Term -> Term -> Formula
  | atomic : forall r : Relations L, Terms (arity L (inl _ r)) -> 
                Formula
  | impH : Formula -> Formula -> Formula
  | notH : Formula -> Formula
  | forallH : nat -> Formula -> Formula.
\end{verbatim}
I defined the other logical connectives in terms of {\code{impH}},
{\code{notH}}, and {\code{forallH}}.

The {\code{H}} at the end of the logic connectives (such as {\code{impH}})
stands for ``Hilbert'' and is used to distinguish them from
{\tmhlink{Coq}{http://coq.inria.fr/}}'s connectives.

For example, the formula $\tmmathbf{\neg} \tmmathbf{\forall}
\tmmathbf{\tmvar{x}}_0 . \tmmathbf{\forall} \tmmathbf{\tmvar{x}}_1 .
\tmmathbf{\tmvar{x}}_0  \tmmathbf{=}  \tmmathbf{\tmvar{x}}_1$ would be
represented by:
\begin{verbatim}
notH (forallH 0 (forallH 1 (equal (var 0) (var 1))))
\end{verbatim}

It would be nice to use higher order abstract syntax to handle bound variables
by giving {\code{forallH}} the type {\code{(Term -> Formula) -> Formula}}. I
would represent the above example as:
\begin{verbatim}
notH (forallH (fun x : Term => 
        (forallH (fun y : Term => (equal x y)))))
\end{verbatim}
This technique would require addition work to disallow ``exotic terms'' that
are created by passing a function into {\code{forallH}} that does a case
analysis on the term and returning entirely different formulas in different
cases. {\tmhlink{Despeyroux et al.}{http://www.site.uottawa.ca/\~{ }afelty/abstracts/tlca95.html}}
{\cite{despeyroux:1994}} address this problem by creating a complicated
predicate that only valid formulas satisfy.

Another choice would have been to use de Bruijn indexes to eliminate named
variables. However dealing with free and bound variables with de Bruijn
indexes can be difficult.

Using named variables allowed me to closely follow
{\tmhlink{Hodel's}{URN:ISBN:053494440X}} work {\cite{hodel:1995}}. Also, in
order to help persuade people that the statement of the incompleteness theorem
is correct, it is helpful to make the underlying definitions as familiar as
possible.

Renaming bound variables turned out to be a constant source of work during
development because variable names and terms were almost always abstract. In
principle the variable names could conflict, so it was constantly necessary to
consider this case and deal with it by renaming a bound variable to a fresh
one. Perhaps it would have been better to use de Bruijn indexes and a
deduction system that only deduced closed formulas.

\subsection{Definition of {\code{substituteFormula}}}

\begin{sloppypar}
I defined the function {\code{substituteFormula}} to substitute a term for
all occurrences of a free variable inside a given formula. While the
definition of {\code{substituteTerm}} is simple structural recursion,
substitution for formulas is complicated by quantifiers. Suppose we want to
substitute the term {\tmvar{s}} for $\tmmathbf{\tmvar{x}}_{\tmvar{i}}$ in the
formula $\tmmathbf{\forall} \tmmathbf{\tmvar{x}}_{\tmvar{j}} .
\tmvar{\varphi}$ and $\tmvar{i} \neq \tmvar{j}$. Suppose
$\tmmathbf{\tmvar{x}}_{\tmvar{j}}$ is a free variable of {\tmvar{s}}. If we
naïvely perform the substitution then the occurrences of
$\tmmathbf{\tmvar{x}}_{\tmvar{j}}$ in {\tmvar{s}} get captured by the
quantifier. One common solution to this problem is to disallow substitution
for a term $\tmvar{s}$ when $\tmvar{s}$ is not {\tmem{substitutable}} for
$\tmvar{x}_{\tmvar{i}}$ in $\tmvar{\varphi}$. The solution I take is to rename
the bound variable in this case.
\[ ( \tmmathbf{\forall} \tmmathbf{\tmvar{x}}_{\tmvar{j}} . \tmvar{\varphi} ) [
   \tmmathbf{\tmvar{x}}_{\tmvar{i}} / \tmvar{s} ] \defeq \tmmathbf{\forall}
   \tmmathbf{\tmvar{x}}_{\tmvar{k}} . ( \tmvar{\varphi} [
   \tmmathbf{\tmvar{x}}_{\tmvar{j}} / \tmmathbf{\tmvar{x}}_{\tmvar{k}} ] ) [
   \tmmathbf{\tmvar{x}}_{\tmvar{i}} / \tmvar{s} ] \tmop{where} \tmvar{k} \neq
   \tmvar{i} \tmop{and} \tmmathbf{\tmvar{x}}_{\tmvar{k}} \tmop{is} \tmop{not}
   \tmop{free} \tmop{in} \tmvar{\varphi} \tmop{or} \tmvar{s} \]
Unfortunately this definition is not structurally recursive. The second
substitution operates on the result of the first substitution, which is not
structurally smaller than the original formula.
\end{sloppypar}
{\tmhlink{Coq}{http://coq.inria.fr/}} will not accept this recursive
definition as is; it is necessary to prove the recursion will terminate. I
proved that substitution preserves the {\tmem{depth}} of a formula, and that
each recursive call operates on a formula of smaller depth.

One of McBride's mantras says, {\tmhlink{``If my recursion is not
structural, I am using the wrong structure''}{http://www.cs.nott.ac.uk/\~{
}ctm/thesis.ps.gz}} {\cite[p. 241]{mcbride:1999}}. In this case, my recursion is
not structural because I am using the wrong recursion. Stoughton shows
that {\tmhlink{it is easier to define substitution that substitutes all
variables simultaneously because the recursion is
structural}{http://www.cis.ksu.edu/\~{ }allen/Stoughton/subst.ps}}
{\cite{stoughton:1988}}. If I had
made this definition first, I could have defined substitution of one variable
in terms of it and many of my difficulties would have disappeared.

\subsection{Definition of {\code{Prf}}}

I defined the inductive type {\code{(Prf Gamma phi)}} to be the type of
proofs of {\code{phi}}, from the list of assumptions {\code{Gamma}}.
\begin{verbatim}
Inductive Prf : Formulas -> Formula -> Set :=
  | AXM : forall A : Formula, Prf (A :: nil) A
  | MP : forall (Axm1 Axm2 : Formulas) (A B : Formula),
      Prf Axm1 (impH A B) -> Prf Axm2 A -> 
         Prf (Axm1 ++ Axm2) B
  | GEN : forall (Axm : Formulas) (A : Formula) (v : nat),
      ~ In v (freeVarListFormula L Axm) -> Prf Axm A -> 
         Prf Axm (forallH v A)
  | IMP1 : forall A B : Formula, Prf nil (impH A (impH B A))
  | IMP2 : forall A B C : Formula,
      Prf nil (impH (impH A (impH B C)) 
                 (impH (impH A B) (impH A C)))
  | CP : forall A B : Formula,
      Prf nil (impH (impH (notH A) (notH B)) (impH B A))
  | FA1 : forall (A : Formula) (v : nat) (t : Term),
      Prf nil (impH (forallH v A) (substituteFormula L A v t))
  | FA2 : forall (A : Formula) (v : nat),
      ~ In v (freeVarFormula L A) -> 
         Prf nil (impH A (forallH v A))
  | FA3 : forall (A B : Formula) (v : nat),
      Prf nil
        (impH (forallH v (impH A B)) 
           (impH (forallH v A) (forallH v B)))
  | EQ1 : Prf nil (equal (var 0) (var 0))
  | EQ2 : Prf nil (impH (equal (var 0) (var 1)) 
                     (equal (var 1) (var 0)))
  | EQ3 : Prf nil
        (impH (equal (var 0) (var 1))
           (impH (equal (var 1) (var 2)) (equal (var 0) (var 2))))
  | EQ4 : forall R : Relations L, Prf nil (AxmEq4 R)
  | EQ5 : forall f : Functions L, Prf nil (AxmEq5 f).
\end{verbatim}
{\code{AxmEq4}} and {\code{AxmEq5}} are recursive functions that generate
the equality axioms for relations and functions. {\code{AxmEq4 R}} generates
\[ \tmmathbf{\tmvar{x}}_0  \tmmathbf{=}  \tmmathbf{\tmvar{x}}_1 
   \tmmathbf{\Rightarrow} \ldots \tmmathbf{\Rightarrow} 
   \tmmathbf{\tmvar{x}}_{2 \tmvar{n} - 2}  \tmmathbf{=} 
   \tmmathbf{\tmvar{x}}_{2 \tmvar{n} - 1}
   \tmmathbf{\Rightarrow}\linebreak ( \tmvar{R}
   ( \tmmathbf{\tmvar{x}}_0, \ldots,
   \tmmathbf{\tmvar{x}}_{2 \tmvar{n} - 2} ) \tmmathbf{\Leftrightarrow} 
   \tmvar{R} ( \tmmathbf{\tmvar{x}}_1, \ldots,
   \tmmathbf{\tmvar{x}}_{2 \tmvar{n} - 1} ) ) \]
and {\code{AxmEq5 f}} generates
\[ \tmmathbf{\tmvar{x}}_0  \tmmathbf{=}  \tmmathbf{\tmvar{x}}_1 
   \tmmathbf{\Rightarrow} \ldots \tmmathbf{\Rightarrow}\linebreak 
   \tmmathbf{\tmvar{x}}_{2 \tmvar{n} - 2}  \tmmathbf{=} 
   \tmmathbf{\tmvar{x}}_{2 \tmvar{n} - 1}  \tmmathbf{\Rightarrow}  \tmvar{f} (
   \tmmathbf{\tmvar{x}}_0, \ldots,
   \tmmathbf{\tmvar{x}}_{2 \tmvar{n} - 2} ) \tmmathbf{=}  \tmvar{f} (
   \tmmathbf{\tmvar{x}}_1, \ldots,
   \tmmathbf{\tmvar{x}}_{2 \tmvar{n} - 1} ) \]

I found that replacing ellipses from informal proofs with recursive functions
was one of the most difficult tasks. The informal proof does not contain
information on what inductive hypothesis should be used when reasoning about
these recursive definitions. Figuring out the correct inductive hypotheses was
not easy.

\subsection{Definition of {\code{SysPrf}}}

There are some problems with the definition of {\code{Prf}} given. It
requires the list of axioms to be in the correct order for the proof. For
example, if we have {\code{Prf Gamma1 (impH phi psi)}} and {\code{Prf Gamma2
phi}} then we can conclude only {\code{Prf Gamma1++Gamma2 psi}}. We cannot
conclude {\code{Prf Gamma2++Gamma1 psi}} or any other permutation of
{\code{psi}}. If an axiom is used more than once, it must appear in the list
more than once. If an axiom is never used, it must not appear. Also, the
number of axioms must be finite because they form a list.

To solve this problem, I defined a {\code{System}} to be {\code{Ensemble
Formula}}, and {\code{(SysPrf T phi)}} to be the proposition that the system
{\code{T}} proves {\code{phi}}.
\begin{verbatim}
Definition System := Ensemble Formula.
Definition mem := Ensembles.In.

Definition SysPrf (T : System) (f : Formula) :=
  exists Axm : Formulas,
    (exists prf : Prf Axm f,
       (forall g : Formula, In g Axm -> mem _ T g)).
\end{verbatim}
{\code{Ensemble A}} represents subsets of {\code{A}} by the functions
{\code{A -> Prop}}. {\code{a : A}} is considered to be a member of {\code{T
: Ensemble A}} if and only if the type {\code{T a}} is inhabited. I also
defined {\code{mem}} to be {\code{Ensembles.In}} so that it does not
conflict with {\code{List.In}}.

\subsection{The Deduction Theorem}

The deduction theorem states that if $\tmvar{\Gamma} \cup \{ \tmvar{\varphi}
\} \vdash \tmvar{\psi}$ then $\tmvar{\Gamma} \vdash \tmvar{\varphi} 
\tmmathbf{\Rightarrow}  \tmvar{\psi}$.

There is a choice of whether the side condition for the
$\forall$-generalization rule, {\code{\~{ } In v (freeVarListFormula L
Axm)}}, should be required or not. If this side condition is removed then the
deduction theorem requires a side condition on it. Usually all the formulas in
an axiom system are closed, so the side condition on the
$\forall$-generalization is easy to show. So I decided to keep the side
condition on the $\forall$-generalization rule.

At one point the proof of the deduction theorem requires proving that if
$\tmvar{\Gamma} \cup \{ \tmvar{\varphi} \} \vdash \tmvar{\psi}$ because
$\tmvar{\psi} \in \tmvar{\Gamma} \cup \{ \tmvar{\varphi} \}$, then
$\tmvar{\Gamma} \vdash \tmvar{\varphi}  \tmmathbf{\Rightarrow}  \tmvar{\psi}$.
There are two cases to consider. If $\tmvar{\psi} = \tmvar{\varphi}$ then the
result easily follows from the reflexivity of $\tmmathbf{\Rightarrow}$.
Otherwise $\tmvar{\psi} \in \tmvar{\Gamma}$, and therefore $\tmvar{\Gamma}
\vdash \tmvar{\psi}$. The result then follows. In order to constructively make
this choice it is necessary to decide whether $\tmvar{\psi} = \tmvar{\varphi}$
or not. This requires {\code{Formula}} to be a decidable type, and that
requires the language {\code{L}} to be decidable. Since {\code{L}} could be
anything, I needed to add hypotheses that the function and relation symbols
are decidable types.
\begin{itemize}
  \item {\code{forall x y : Functions L,  \{ x=y \}  +  \{ x<>y \}}}
  
  \item {\code{forall x y : Relations L,  \{ x=y \}  +  \{ x<>y \}}}.
\end{itemize}

I used the deduction theorem without restriction and ended up using the
hypotheses in many lemmas. I expect that many of these lemmas could be proved
without assuming the decidability of the language. It is hard to imagine a
useful language that is not decidable, so I do not feel too bad about using
these hypotheses in unnecessary places.

\subsection{Languages and Theories of Number Theory}

I created two languages. The first language, {\code{LNT}}, is the language of
number theory and just has the function symbols {\code{Plus}},
{\code{Times}}, {\code{Succ}}, and {\code{Zero}} with appropriate arities.
The second language, {\code{LNN}}, is the language of NN and has the same
function symbols as {\code{LNT}} plus one relation symbol for less than,
{\code{LT}}.

\label{AXIOMSYSTEMS}I define two axiom systems: {\code{NN}} and {\code{PA}}.
{\code{NN}} and {\code{PA}} share six axioms.
\begin{enumerate}
  \item $\tmmathbf{\forall} \tmmathbf{\tmvar{x}}_0 . \tmmathbf{\neg} 
  \tmmathbf{S} \tmmathbf{\tmvar{x}}_0  \tmmathbf{=}  \tmmathbf{0}$
  
  \item $\tmmathbf{\forall} \tmmathbf{\tmvar{x}}_0 . \tmmathbf{\forall}
  \tmmathbf{\tmvar{x}}_1 . ( \tmmathbf{S} \tmmathbf{\tmvar{x}}_0  \tmmathbf{=}
  \tmmathbf{S} \tmmathbf{\tmvar{x}}_1  \tmmathbf{\Rightarrow} 
  \tmmathbf{\tmvar{x}}_0  \tmmathbf{=}  \tmmathbf{\tmvar{x}}_1 )$
  
  \item $\tmmathbf{\forall} \tmmathbf{\tmvar{x}}_0 . \tmmathbf{\tmvar{x}}_0 
  \tmmathbf{+}  \tmmathbf{0}  \tmmathbf{=}  \tmmathbf{\tmvar{x}}_0$
  
  \item $\tmmathbf{\forall} \tmmathbf{\tmvar{x}}_0 . \tmmathbf{\forall}
  \tmmathbf{\tmvar{x}}_1 . \tmmathbf{\tmvar{x}}_0  \tmmathbf{+}  \tmmathbf{S}
  \tmmathbf{\tmvar{x}}_1  \tmmathbf{=}  \tmmathbf{S} ( \tmmathbf{\tmvar{x}}_0 
  \tmmathbf{+}  \tmmathbf{\tmvar{x}}_1 )$
  
  \item $\tmmathbf{\forall} \tmmathbf{\tmvar{x}}_0 . \tmmathbf{\tmvar{x}}_0 
  \tmmathbf{\times}  \tmmathbf{0}  \tmmathbf{=}  \tmmathbf{0}$
  
  \item $\tmmathbf{\forall} \tmmathbf{\tmvar{x}}_0 . \tmmathbf{\forall}
  \tmmathbf{\tmvar{x}}_1 . \tmmathbf{\tmvar{x}}_0  \tmmathbf{\times} 
  \tmmathbf{S} \tmmathbf{\tmvar{x}}_1  \tmmathbf{=} ( \tmmathbf{\tmvar{x}}_0 
  \tmmathbf{\times}  \tmmathbf{\tmvar{x}}_1 ) \tmmathbf{+} 
  \tmmathbf{\tmvar{x}}_0$
\end{enumerate}
{\code{NN}} has three additional axioms about less than.
\begin{enumerate}
  \item $\tmmathbf{\forall} \tmmathbf{\tmvar{x}}_0 . \tmmathbf{\neg}
  \tmmathbf{\tmvar{x}}_0  \tmmathbf{<}  \tmmathbf{0}$
  
  \item $\tmmathbf{\forall} \tmmathbf{\tmvar{x}}_0 . \tmmathbf{\forall}
  \tmmathbf{\tmvar{x}}_1 . ( \tmmathbf{\tmvar{x}}_0  \tmmathbf{<} 
  \tmmathbf{S} \tmmathbf{\tmvar{x}}_1  \tmmathbf{\Rightarrow} (
  \tmmathbf{\tmvar{x}}_0  \tmmathbf{=}  \tmmathbf{\tmvar{x}}_1 
  \tmmathbf{\vee}  \tmmathbf{\tmvar{x}}_0  \tmmathbf{<} 
  \tmmathbf{\tmvar{x}}_1 ) )$
  
  \item $\tmmathbf{\forall} \tmmathbf{\tmvar{x}}_0 . \tmmathbf{\forall}
  \tmmathbf{\tmvar{x}}_1 . ( \tmmathbf{\tmvar{x}}_0  \tmmathbf{<} 
  \tmmathbf{\tmvar{x}}_1  \tmmathbf{\vee}  \tmmathbf{\tmvar{x}}_0 
  \tmmathbf{=}  \tmmathbf{\tmvar{x}}_1  \tmmathbf{\vee} 
  \tmmathbf{\tmvar{x}}_1  \tmmathbf{<}  \tmmathbf{\tmvar{x}}_0 )$
\end{enumerate}
{\code{PA}} has an infinite number of induction axioms that follow one
schema.
\begin{enumerate}
  \item (schema) $\tmmathbf{\forall} \tmmathbf{\tmvar{x}}_{\tmvar{i}_1} .
  \ldots \tmmathbf{\forall} \tmmathbf{\tmvar{x}}_{\tmvar{i}_{\tmvar{n}}} .
  \tmvar{\varphi} [ \tmmathbf{\tmvar{x}}_{\tmvar{j}} / \tmmathbf{0} ]
  \tmmathbf{\Rightarrow}  \tmmathbf{\forall} \tmmathbf{\tmvar{x}}_{\tmvar{j}}
  . ( \tmvar{\varphi}  \tmmathbf{\Rightarrow}  \tmvar{\varphi} [
  \tmmathbf{\tmvar{x}}_{\tmvar{j}} / \tmmathbf{S}
  \tmmathbf{\tmvar{x}}_{\tmvar{j}} ] ) \tmmathbf{\Rightarrow} 
  \tmmathbf{\forall} \tmmathbf{\tmvar{x}}_{\tmvar{j}} . \tmvar{\varphi}$
\end{enumerate}
The $\tmmathbf{\tmvar{x}}_{\tmvar{i}_1}, \ldots,
\tmmathbf{\tmvar{x}}_{\tmvar{i}_{\tmvar{n}}}$ are the free variables of
$\tmmathbf{\forall} \tmmathbf{\tmvar{x}}_{\tmvar{j}} . \tmvar{\varphi}$. The
quantifiers ensure that all the axioms of {\tmabbr{PA}} are closed.

Because NN is in a different language than {\tmabbr{PA}}, a proof in NN is not
a proof in {\tmabbr{PA}}. In order to reuse the work done in NN, I created a
function called {\code{LNN2LNT\_formula}} to convert formulas in
{\code{LNN}} into formulas in {\code{LNT}} by replacing occurrences of
$\tmvar{t}_0  \tmmathbf{<}  \tmvar{t}_1$ with $( \exists
\tmmathbf{\tmvar{x}}_2 . \tmmathbf{\tmvar{x}}_0 + ( \tmmathbf{S}
\tmmathbf{\tmvar{x}}_2 ) = \tmmathbf{\tmvar{x}}_1 ) [ \tmmathbf{\tmvar{x}}_0 /
\tmvar{t}_0, \tmmathbf{\tmvar{x}}_1 / \tmvar{t}_1 ]$---$\tmvar{\varphi} [
\tmmathbf{\tmvar{x}}_0 / \tmvar{t}_0, \tmmathbf{\tmvar{x}}_1 / \tmvar{t}_1 ]$
is the simultaneous substitution of $\tmvar{t}_0$ for $\tmmathbf{\tmvar{x}}_0$
and $\tmvar{t}_1$ for $\tmmathbf{\tmvar{x}}_1$. Then I proved that if
$\tmop{NN} \vdash \tmvar{\varphi}$ then $\tmabbr{\tmop{PA}} \vdash
\text{{\code{LNN2LNT\_formula}}} ( \tmvar{\varphi} )$.

I also created the function {\code{natToTerm : nat -> Term}} to return the
closed term representing a given natural number. In this document I will refer
to this function as $\ulcorner \centerdot \urcorner$, so $\ulcorner 0
\urcorner = \tmmathbf{0}$, $\ulcorner 1 \urcorner = \tmmathbf{S}
\tmmathbf{0}$, etc.

\section{Coding} \label{C}

To prove the incompleteness theorem, it is necessary for the inner logic to
reason about proofs and formulas, but the inner logic can only reason about
natural numbers. It is therefore necessary to code proofs and formulas as
natural numbers.

Gödel's original approach was to code a formula as a list of numbers and then
code that list using properties from the prime decomposition
theorem{\cite{godel:1931}}. I avoided needing theorems about prime
decomposition by using the Cantor pairing function instead. The Cantor pairing
function, {\code{cPair}}, is a commonly used bijection between $\mathbbm{N}
\times \mathbbm{N}$ and $\mathbbm{N}$.
\[ \text{{\code{cPair}}} ( \tmvar{a}, \tmvar{b} ) \defeq \tmvar{a} + \sum_{i
   = 1}^{a + b}  \tmvar{i} \]
All my inductive structures were easy to recursively encode. I gave each
constructor a unique number and paired that number with the encoding of all
its parameters. For example, I defined {\code{codeFormula}} as:
\begin{verbatim}
Fixpoint codeFormula (f : Formula) : nat :=
  match f with
  | fol.equal t1 t2 => cPair 0 (cPair (codeTerm t1) (codeTerm t2))
  | fol.impH f1 f2 => 
       cPair 1 (cPair (codeFormula f1) (codeFormula f2))
  | fol.notH f1 => cPair 2 (codeFormula f1)
  | fol.forallH n f1 => cPair 3 (cPair n (codeFormula f1))
  | fol.atomic R ts => cPair (4+(codeR R)) (codeTerms _ ts)
  end.
\end{verbatim}
where {\code{codeR}} is a coding of the relation symbols for the language.

I will use $\ulcorner \tmvar{\varphi} \urcorner$ for $\ulcorner
\tmop{{\code{codeFormula}}} \; \tmvar{\varphi} \urcorner$ and $\ulcorner
\tmvar{t} \urcorner$ for $\ulcorner \tmop{{\code{codeTerm}}} \; \tmvar{t}
\urcorner$.

\section{The Statement of Incompleteness} \label{TSI}

The incompleteness theorem states the essential incompleteness of NN, meaning
that for every axiom system {\tmvar{T}} such that
\begin{itemize}
  \item $\tmop{NN} \subseteq T$
  
  \item {\tmvar{T}} can {\tmhlink{represent its own axioms}{\#
  REPRESENTSINSELF}}
  
  \item {\tmvar{T}} is a {\tmhlink{decidable set}{\# DECIDABLESET}}
\end{itemize}
then there exists a sentence $\varphi$ such that if $\tmvar{T} \vdash
\tmvar{\varphi}$ or $\tmvar{T} \vdash \tmmathbf{\neg} \tmvar{\varphi}$ then
{\tmvar{T}} is {\tmhlink{inconsistent}{\# INCONSISTENT}}.

The theorem is only about proofs in {\code{LNN}}, the language of NN. This
statement does not show the incompleteness of theories that extend the
language.

In {\tmhlink{Coq}{http://coq.inria.fr/}} the theorem is stated as as:
\begin{verbatim}
Theorem Incompleteness
     : forall T : System,
       Included Formula NN T ->
       RepresentsInSelf T ->
       DecidableSet Formula T ->
       exists f : Formula,
         Sentence f /\
         (SysPrf T f \/ SysPrf T (notH f) -> Inconsistent LNN T).
\end{verbatim}
\label{INCONSISTENT}A {\code{System}} is {\code{Inconsistent}} if it proves
all formulas.
\begin{verbatim}
Definition Inconsistent (T : System) := 
   forall f : Formula, SysPrf T f.
\end{verbatim}

A {\code{Sentence}} is a {\code{Formula}} without any free variables.
\begin{verbatim}
Definition Sentence (f : Formula) := 
   forall v : nat, ~ In v (freeVarFormula LNN f).
\end{verbatim}

\label{DECIDABLESET}A {\code{DecidableSet}} is an {\code{Ensemble}} such
that every item either belongs to the {\code{Ensemble}} or does not belong to
the {\code{Ensemble}}. This hypothesis is trivially true in classical logic,
but in constructive logic I needed it to prove the strong constructive existential
quantifier in the statement of incompleteness.
\begin{verbatim}
Definition DecidableSet (A : Type)(s : Ensemble A) :=
   forall x : A, mem A s x \/ ~ mem A s x.
\end{verbatim}

\label{REPRESENTSINSELF}The {\code{RepresentsInSelf}} hypothesis restricts
what the {\code{System}} {\tmvar{T}} can be. The statement of essential
incompleteness normally requires {\tmvar{T}} be a recursive set. Instead I use
the weaker hypothesis that the set {\tmvar{T}} is expressible in the system
{\tmvar{T}}.

\label{EXPRESSIBLE}Given a system $\tmvar{T}$ extending NN and another system
$\tmvar{U}$ along with a formula $\tmvar{\varphi}_{\tmvar{U}}$ with at most
one free variable $\tmmathbf{\tmvar{x}}_{\tmvar{i}}$, we say
$\tmvar{\varphi}_{\tmvar{U}}$ expresses the axiom system $\tmvar{U}$ in
$\tmvar{T}$ if the following hold for all formulas $\tmvar{\psi}$.
\begin{enumerate}
  \item if $\tmvar{\psi} \in \tmvar{U}$ then $\tmvar{T} \vdash
  \tmvar{\varphi}_{\tmvar{U}} [ \tmmathbf{\tmvar{x}}_{\tmvar{i}} / \ulcorner
  \tmvar{\psi} \urcorner ]$
  
  \item if $\tmvar{\psi} \nin \tmvar{U}$ then $\tmvar{T} \vdash
  \tmmathbf{\neg} \tmvar{\varphi}_{\tmvar{U}} [
  \tmmathbf{\tmvar{x}}_{\tmvar{i}} / \ulcorner \tmvar{\psi} \urcorner ]$
\end{enumerate}
$\tmvar{U}$ is expressible in $\tmvar{T}$ if there exists a formula
$\tmvar{\varphi}_{\tmvar{U}}$ such that $\tmvar{\varphi}_{\tmvar{U}}$
expresses the axiom system $\tmvar{U}$ in $\tmvar{T}$.

In {\tmhlink{Coq}{http://coq.inria.fr/}} I write the statement $\tmvar{T}$ is
expressible in $\tmvar{T}$ as
\begin{verbatim}
Definition RepresentsInSelf (T : System) := 
exists rep : Formula, exists v : nat,
(forall x : nat, In x (freeVarFormula LNN rep) -> x = v)  /\
(forall f : Formula,
        mem Formula T f ->
        SysPrf T (substituteFormula LNN rep v 
                    (natToTerm (codeFormula f)))) /\
(forall f : Formula,
        ~ mem Formula T f ->
        SysPrf T (notH (substituteFormula LNN rep v 
                          (natToTerm (codeFormula f))))).
\end{verbatim}
This is weaker than requiring that {\tmvar{T}} be a recursive set because any
recursive set of axioms {\tmvar{T}} is expressible in NN. Since {\tmvar{T}} is
an extension of NN, any recursive set of axioms {\tmvar{T}} is expressible in
{\tmvar{T}}.

By using this weaker hypothesis I avoid defining what a recursive set is.
Also, in this form the theorem could be used to prove that any complete and
consistent theory of arithmetic cannot define its own axioms. In particular,
this could be used to prove Tarski's theorem that the truth predicate is not
definable.

\section{Primitive Recursive Functions} \label{PRF}

A common approach to proving the incompleteness theorem is to prove that every
primitive recursive function is representable. Informally an {\tmvar{n}}-ary
function {\tmvar{f}} is representable in NN if there exists a formula
$\tmvar{\varphi}$ such that
\begin{enumerate}
  \item the free variables of $\tmvar{\varphi}$ are among
  $\tmmathbf{\tmvar{x}}_0, \ldots, \tmmathbf{\tmvar{x}}_{\tmvar{n}}$.
  
  \item for all $\tmvar{a}_1, \ldots, \tmvar{a}_{\tmvar{n}} :\mathbbm{N}$,\\
  $\tmop{NN} \vdash ( \tmvar{\varphi}  \tmmathbf{\Rightarrow} 
  \tmmathbf{\tmvar{x}}_0  \tmmathbf{=} \ulcorner \tmvar{f} ( \tmvar{a}_1,
  \ldots, \tmvar{a}_{\tmvar{n}} ) \urcorner ) [ \tmmathbf{\tmvar{x}}_1 /
  \ulcorner \tmvar{a}_1 \urcorner, \ldots, \tmmathbf{\tmvar{x}}_{\tmvar{n}} /
  \ulcorner \tmvar{a}_{\tmvar{n}} \urcorner ]$
\end{enumerate}
I defined the type {\code{PrimRec n}} as:
\begin{verbatim}
Inductive PrimRec : nat -> Set :=
  | succFunc : PrimRec 1
  | zeroFunc : PrimRec 0
  | projFunc : forall n m : nat, m < n -> PrimRec n
  | composeFunc :
      forall (n m : nat) (g : PrimRecs n m) (h : PrimRec m), 
         PrimRec n
  | primRecFunc :
      forall (n : nat) (g : PrimRec n) (h : PrimRec (S (S n))), 
         PrimRec (S n)
with PrimRecs : nat -> nat -> Set :=
  | PRnil : forall n : nat, PrimRecs n 0
  | PRcons : forall n m : nat, 
      PrimRec n -> PrimRecs n m -> PrimRecs n (S m).
\end{verbatim}
{\code{PrimRec n}} is the expression of an {\tmvar{n}}-ary primitive
recursive function, but it is not itself a function. I defined
{\code{evalPrimRec : forall n : nat, PrimRec n -> naryFunc n}} to convert the
expression into a function. Rather than working directly with primitive
recursive expressions, I worked with particular
{\tmhlink{Coq}{http://coq.inria.fr/}} functions and proved they were
extensionally equivalent to the evaluation of primitive recursive expressions.

I proved that every primitive recursive function is representable in NN. This
required using {\tmhlink{Gödel's
$\beta$-function}{http://www.math.uwaterloo.ca/\~{ }snburris/htdocs/scav/fo\_arith/fo\_arith.html}} along with the Chinese remainder theorem. The
$\beta$-function is a function that codes array indexing. A finite list of
numbers $\tmvar{a}_0, \ldots, \tmvar{a}_{\tmvar{n}}$ is coded as a pair of
numbers $( \tmvar{x}, \tmvar{y} )$ and $\beta ( \tmvar{x}, \tmvar{y},
\tmvar{i} ) = \tmvar{a}_{\tmvar{i}}$. The $\beta$-function is special because
it is defined in terms of plus and times and is non-recursive. The Chinese
remainder theorem is used to prove that the $\beta$-function works.

I took care to make the formulas representing the primitive recursive
functions clearly $\Sigma_1$ by ensuring that only the unbounded quantifiers
are existential; however, I did not prove that the formulas are $\Sigma_1$
because it is not needed for the first incompleteness theorem. Such a proof
could be used for the second incompleteness theorem {\cite{shoenfield:1967}}.

\subsection{{\code{codeSubFormula}} is Primitive Recursive}

I proved that substitution is primitive recursive. Since substitution is
defined in terms of {\code{Formula}} and {\code{Term}}, it itself cannot be
primitive recursive. Instead I proved that the corresponding function
operating on codes is primitive recursive. This function is called
{\code{codeSubFormula}} and I proved it is correct in the following sense.
\[ \text{{\code{codeSubFormula}}} ( \ulcorner \tmvar{\varphi} \urcorner,
   \tmvar{i}, \ulcorner \tmvar{s} \urcorner ) = \ulcorner \tmvar{\varphi} [
   \tmmathbf{\tmvar{x}}_{\tmvar{i}} / \tmvar{s} ] \urcorner \]

Next I proved that it is primitive recursive. This proof is very difficult.
The problem is again with the need to rebind bound variables. Normally one
would attempt to create this primitive recursive function by using
course-of-values recursion. Course-of-values recursion requires all recursive
calls have a smaller code than the original call. Renaming a bound variable
requires two recursive calls. Recall the definition of substitution in this
case:
\[ ( \tmmathbf{\forall} \tmmathbf{\tmvar{x}}_{\tmvar{j}} . \tmvar{\varphi} ) [
   \tmmathbf{\tmvar{x}}_{\tmvar{i}} / \tmvar{s} ] \defeq \tmmathbf{\forall}
   \tmmathbf{\tmvar{x}}_{\tmvar{k}} . ( \tmvar{\varphi} [
   \tmmathbf{\tmvar{x}}_{\tmvar{j}} / \tmmathbf{\tmvar{x}}_{\tmvar{k}} ] ) [
   \tmmathbf{\tmvar{x}}_{\tmvar{i}} / \tmvar{s} ] \tmop{where} \tmvar{k} \neq
   \tmvar{i} \tmop{and} \tmmathbf{\tmvar{x}}_{\tmvar{k}} \tmop{is} \tmop{not}
   \tmop{free} \tmop{in} \tmvar{\varphi} \tmop{or} \tmvar{s} \]
If one is lucky one might be able to make the inner recursive call. But there
is no reason to suspect the input to the second recursive call,
$\tmvar{\varphi} [ \tmmathbf{\tmvar{x}}_{\tmvar{j}} /
\tmmathbf{\tmvar{x}}_{\tmvar{k}} ]$, is going to have a smaller code than the
original input, $\tmmathbf{\forall} \tmmathbf{\tmvar{x}}_{\tmvar{j}} .
\tmvar{\varphi}$.

If I had used the alternative definition of substitution, where all variables
are substituted simultaneously, there would still be problems. The input would
include a list of variable and term pairs. In this case a new pair would be
added to the list when making the recursive call, so the input to the
recursive call could still have a larger code than the input to the original
call.

It seems that using course-of-values recursion is difficult or impossible.
Instead I introduce the notion of the trace of the computation of
substitution. Think of the trace of computation as a finite tree where the
nodes contain the input and output of each recursive call. The subtrees of a
node are the traces of the computation of the recursive calls. This tree can
be coded as a number. I proved that there is a primitive recursive function
that can check to see if a number represents a trace of the computation of
substitution.

The key to solving this problem is to create a primitive recursive function
that computes a bound on how large the code of the trace of computation can be
for a given input. With this I created another primitive recursive function
that searches for the trace of computation up to this bound. Once the trace is
found---I proved that it must be found---the function extracts the result from
the trace and returns it.

\subsection{{\code{checkPrf}} is Primitive Recursive}

Given a code for a formula and a code for a proof, the function
{\code{checkPrf}} returns 0 if the proof does not prove the formula,
otherwise it returns one plus the code of the list of axioms used in the
proof. I proved this function is primitive recursive, as well as proving that
it is correct in the sense that for every proof {\tmvar{p}} of
$\tmvar{\varphi}$ from a list of axioms $\tmvar{\Gamma}$,
$\text{{\code{checkPrf}}} ( \ulcorner \tmvar{\varphi} \urcorner, \ulcorner
\tmvar{p} \urcorner ) = 1 + \ulcorner \tmvar{\Gamma} \urcorner ;$ and for all
$\tmvar{n}, \tmvar{m} :\mathbbm{N}$ if $\text{{\code{checkPrf}}} ( \tmvar{n},
\tmvar{m} ) \neq 0$ then there exists $\tmvar{\varphi}$, $\tmvar{\Gamma}$, and
some proof {\tmvar{p}} of $\tmvar{\varphi}$ from $\tmvar{\Gamma}$ such that
$\ulcorner \tmvar{\varphi} \urcorner = \tmvar{n}$ and $\ulcorner \tmvar{p}
\urcorner = \tmvar{m}$.

For any {\tmhlink{axiom system {\tmvar{U}} expressible in {\tmvar{T}}}{\#
EXPRESSIBLE}}, I created the formulas {\code{codeSysPrf}} and
{\code{codeSysPf}}. $\text{{\code{codeSysPrf}}} [ \tmmathbf{\tmvar{x}}_0 /
\ulcorner \tmvar{n} \urcorner, \tmmathbf{\tmvar{x}}_1 / \ulcorner \tmvar{m}
\urcorner ]$ is provable in {\tmvar{T}} if $\tmvar{m}$ is the code of a proof
in {\tmvar{U}} of a formula coded by $\tmvar{n}$. $\text{{\code{codeSysPf}}}
[ \tmmathbf{\tmvar{x}}_0 / \ulcorner \tmvar{n} \urcorner ]$ is provable in
{\tmvar{T}} if there exists a proof in {\tmvar{U}} of a formula coded by
$\tmvar{n}$.

{\code{codeSysPrf}} and {\code{codeSysPf}} are not derived from a primitive
recursive functions because I wanted to prove the incompleteness of axiom
systems that may not have a primitive recursive characteristic function.

\section{Fixed Point Theorem and Rosser's Incompleteness Theorem}
\label{FPTRIT}

The fixed point theorem states that for every formula $\tmvar{\varphi}$ there
is some formula $\tmvar{\psi}$ such that
\[ \tmop{NN} \vdash \tmvar{\psi}  \tmmathbf{\Leftrightarrow}  \tmvar{\varphi}
   [ \tmmathbf{\tmvar{x}}_{\tmvar{i}} / \ulcorner \tmvar{\psi} \urcorner ] \]
and that the free variables of $\tmvar{\psi}$ are that of $\tmvar{\varphi}$
less $\tmmathbf{\tmvar{x}}_{\tmvar{i}}$.

The fixed point theorem allows one to create ``self-referential sentences''. I
used this to create Rosser's sentence which states that for every code of a
proof of itself, there is a smaller code of a proof of its negation. The proof
of Rosser's incompleteness theorem requires doing a bounded search for a
proof, and this requires knowing what is and what is not a proof in the
system. For this reason, I require the decidability of the axiom system.
Without a decision procedure for the axiom system, I cannot constructively do
the search.

\subsection{Incompleteness of {\tmabbr{PA}}}

To demonstrate the incompleteness theorem I used it to prove the
incompleteness of {\tmabbr{PA}}. I created a primitive recursive predicate for
the codes of the axioms of {\tmabbr{PA}}.
{\tmhlink{Coq}{http://coq.inria.fr/}} is sufficiently powerful to prove the
consistency of {\tmabbr{PA}} by proving that the natural numbers model
{\tmabbr{PA}}.

One subtle point is that {\tmhlink{Coq}{http://coq.inria.fr/}}'s logic is
constructive while the internal logic is classical. One cannot interpret a
formula of the internal logic directly in
{\tmhlink{Coq}{http://coq.inria.fr/}} and expect it to be provable if it is
provable in the internal logic. Instead I use a double negation translation of
the formulas. The translated formula will always hold if it holds in the
internal logic.

The consistency of {\tmabbr{PA}} along with the expressibility of its axioms
and the translations of proofs from NN to {\tmabbr{PA}} allowed me to apply
Rosser's incompleteness theorem and prove the incompleteness of
{\tmabbr{PA}}---there exists a sentence $\varphi$ such that neither $\tmop{PA}
\vdash \tmvar{\varphi}$ nor $\tmop{PA} \vdash \tmmathbf{\neg}
\tmvar{\varphi}$.
\begin{verbatim}
Theorem PAIncomplete :
 exists f : Formula,
   (forall v : nat, ~ In v (freeVarFormula LNT f)) /\
   ~ (SysPrf PA f \/ SysPrf PA (notH f)).
\end{verbatim}
\section{Remarks} \label{R}

\subsection{Extracting the Sentence}

Because my proof is constructive, it is possible, in principle, to compute
this sentence that makes {\tmabbr{PA}} incomplete. This was not done for two
reasons. The first reason is that the existential statement lives in
{\tmhlink{Coq}{http://coq.inria.fr/}}'s {\code{Prop}} universe, and
{\tmhlink{Coq}{http://coq.inria.fr/}}'s only extracts from its {\code{Set}}
universe. This was an error on my part. I should have used
{\tmhlink{Coq}{http://coq.inria.fr/}}'s {\code{Set}} existential quantifier;
this problem would be fairly easy to fix. The second reason is that the
sentence contains a closed term of the code of most of itself. I believe this
code is a very large number and it is written in unary notation. This would
likely make the sentence far to large to be actually printed.

\subsection{Robinson's System Q}

The proof of essential incompleteness is usually carried out for Robinson's
system Q. Instead I followed {\tmhlink{Hodel's}{URN:ISBN:053494440X}}
development {\cite{hodel:1995}} and used NN. Q is {\tmabbr{PA}} with the
induction schema replaced with $\tmmathbf{\forall} \tmmathbf{\tmvar{x}}_0 .
\tmmathbf{\exists} \tmmathbf{\tmvar{x}}_1 . ( \tmmathbf{\tmvar{x}}_0
\tmmathbf{=} \tmmathbf{0}  \tmmathbf{\vee}  \tmmathbf{\tmvar{x}}_0
\tmmathbf{=} \tmmathbf{S} \tmmathbf{\tmvar{x}}_1 )$. All of NN axioms are
$\Pi_1$ whereas Q has the above $\Pi_2$ axiom. Both axiom systems are finite.

Neither system is strictly weaker than the other, so it would not be possible
to use the essential incompleteness of one to get the essential incompleteness
of the other; however both NN and Q are sufficiently powerful to prove a small
number of needed lemmas, and afterward only these lemmas are used. If one
abstracts my proof at these lemmas, it would then be easy to prove the
essential incompleteness of both Q and NN.

\subsection{Comparisons with Shankar's 1986 Proof}

It is worth noting the differences between this formalization of the
incompleteness theorem and Shankar's 1986 proof in the Boyer-Moore theorem
prover. The most notable difference is the proof systems. In
{\tmhlink{Coq}{http://coq.inria.fr/}} the user is expected to input the proof,
in the form of a proof script, and {\tmhlink{Coq}{http://coq.inria.fr/}} will
check the correctness of the proof. In the Boyer-Moore theorem prover the user
states a series of lemmas and the system generates the proofs. However, using
the Boyer-Moore proof system requires feeding it a ``well-chosen sequence of
lemmas'' {\cite[p. xii]{shankar:1994}}, so it
would seem the information being fed into the two systems is similar.

There are some notable semantic differences between Shankar's statement of
incompleteness and mine. His theorem only states that finite extensions of Z2,
hereditarily finite set theory, are incomplete, whereas my theorem states that
even infinite extensions of NN are incomplete as long as they are
self-representable. Also Shankar's internal logic allows axioms to define new
relation or function symbols as long as they come with the required proofs of
admissibility. Such extensions are conservative over Z2, but no computer
verified proof of this fact is given. My internal logic does not allow new
symbols. Finally, I prove the essential incompleteness of NN, which is in the
language of arithmetic. Without any set structures the proof is somewhat more
difficult because it requires using Gödel's $\beta$-function.

One of Shankar's goals when creating his proof was to use a proof system
without modifications. Unfortunately he was not able to meet that goal; he
ended up making some improvements to the Boyer-Moore theorem prover. My proof
was developed in {\tmhlink{Coq}{http://coq.inria.fr/}} without any
modifications.

\subsection{Gödel's Second Incompleteness Theorem}

The second incompleteness theorem states that if {\tmvar{T}} is a recursive
system extending {\tmabbr{PA}}---actually a weaker system could be used
here---and $\tmvar{T} \vdash \tmop{Con}_{\tmvar{T}}$ then {\tmvar{T}} is
{\tmhlink{inconsistent}{\# INCONSISTENT}}. $\tmop{Con}_{\tmvar{T}}$ is some
reasonable formula stating the consistency of {\tmvar{T}}, such as
$\tmmathbf{\neg} \Pr_{\tmvar{T}} ( \ulcorner \tmmathbf{0}  \tmmathbf{=} 
\tmmathbf{S} \tmmathbf{0} \urcorner )$, where $\Pr_{\tmvar{T}}$ is the
provability predicate {\code{codeSysPf}} for {\tmvar{T}}.

\begin{sloppypar}
If I had created a formal proof in {\tmabbr{PA}}, I would have
$\vdash_{\tmop{PA}}~\text{``Gödel's first incompleteness theorem''}$. This
could then be mechanically transformed to create another formal proof in
{\tmabbr{PA}} that $\vdash_{\tmop{PA}} ( \tmop{PA} \vdash \text{``Gödel's
first incompleteness theorem''} )$. The reader can verify that the
second incompleteness theorem follows from this. Unfortunately I have only shown that
$\vdash_{\tmhlink{Coq}{http://coq.inria.fr/}} \text{``Gödel's first incompleteness theorem''}$, so the above
argument cannot be used to create a proof of the second incompleteness
theorem.
\end{sloppypar}

Still, this work can be used as a basis for formalizing the second
incompleteness theorem. The approach would be to formalize the
Hilbert-Bernays-Löb derivability conditions:
\begin{enumerate}
  \item if $\tmop{PA} \vdash \tmvar{\varphi}$ then $\tmop{PA} \vdash
  \Pr_{\tmop{PA}} ( \ulcorner \tmvar{\varphi} \urcorner )$
  
  \item $\tmop{PA} \vdash \Pr_{\tmop{PA}} ( \ulcorner \tmvar{\varphi}
  \urcorner ) \tmmathbf{\Rightarrow} \Pr_{\tmop{PA}} ( \ulcorner
  \Pr_{\tmop{PA}} ( \ulcorner \tmvar{\varphi} \urcorner ) \urcorner )$
  
  \item $\tmop{PA} \vdash \Pr_{\tmop{PA}} ( \ulcorner \tmvar{\varphi} 
  \tmmathbf{\Rightarrow}  \tmvar{\psi} \urcorner ) \tmmathbf{\Rightarrow}
  \Pr_{\tmop{PA}} ( \ulcorner \tmvar{\varphi} \urcorner )
  \tmmathbf{\Rightarrow} \Pr_{\tmop{PA}} ( \ulcorner \tmvar{\psi} \urcorner )$
\end{enumerate}
The second condition is the most difficult to prove. It is usually proved by
first proving that for every $\Sigma_1$ sentence $\tmvar{\varphi}$, $\tmop{PA}
\vdash \tmvar{\varphi}  \tmmathbf{\Rightarrow} \Pr_{\tmop{PA}} ( \ulcorner
\tmvar{\varphi} \urcorner )$. Because I made sure that all primitive recursive
functions are representable by a $\Sigma_1$ formula, it would be easy to go
from this theorem to the second Hilbert-Bernays-Löb condition.

\section{Statistics} \label{S}

My proof, excluding standard libraries and the {\tmhlink{library for
Pocklington's criterion}{http://coq.inria.fr/contribs/pocklington.html}}
{\cite{oostdijk:2001}}, consists of 46 source files, 7 036 lines of
specifications, 37 906 lines of proof, and 1 267 747 total characters. The
size of the gzipped tarball ({\tmkbd{gzip -9}}) of all the source files is
146 008 bytes, which is an estimate of the information content of my proof.

\section{Acknowledgements} \label{A}

I would like to thank {\tmhlink{{\tmabbr{NSERC}}}{http://www.nserc.ca/}} for
providing funding for this research. I thank {\tmhlink{Robert
Schneck}{http://math.berkeley.edu/\~{ }schneck/}} for introducing me to
{\tmhlink{Coq}{http://coq.inria.fr/}}, and helping me out at the beginning. I
would like to thank {\tmhlink{Nikita Borisov}{http://nikita.ca/}} for letting
me use his computer when the proof became to large for my poor laptop. I would
also like to thank my Berkeley advisor, Leo Harrington, for his advice on
improving upon Hodel's proof. And last, but not least, thanks to the
{\tmhlink{Coq}{http://coq.inria.fr/}} team, because without
{\tmhlink{Coq}{http://coq.inria.fr/}} there would be no proof.

\bibliographystyle{plain} \bibliography{godel.bib}

\end{document}